\documentclass[prd,twocolumn,showpacs,floatfix,amsmath,nofootinbib,amssymb,floatfix]{revtex4}
\usepackage{graphicx,color,dcolumn,booktabs,bm,multirow}
\usepackage{longtable,lscape}
\usepackage{txfonts}
\usepackage{overpic}
\usepackage{amssymb}
\usepackage{array}
\usepackage{indentfirst}
\usepackage{feynmf}   
\usepackage{slashed}  
\usepackage{cases}
\usepackage{color}
\usepackage{multirow}
\usepackage{epstopdf}
\usepackage{graphicx,color,dcolumn,booktabs,bm}
\usepackage[colorlinks,
            citecolor=blue,
            anchorcolor=red,
            menucolor=red,
            linkcolor=red,
            filecolor=red,
            runcolor=red,
            urlcolor=blue,
            frenchlinks=red]{hyperref}

\def\k{\mathbf{k}}
\begin{document}

\title{$D$ wave charmonia}
\author{Dian-Yong Chen$^{1}$\footnote{Corresponding author}}\email{chendy@seu.edu.cn}
\author{Cheng-Qun Pang$^{2}$}\email{pcq@qhnu.edu.cn}
\author{Jun He$^{3}$}\email{junhe@njnu.edu.cn}
\author{Zhi-Yong Zhou$^{1}$}\email{zhouzhy@seu.edu.cn}
\affiliation{$^1$ School of Physics, Southeast University, Nanjing 210094, People's Republic of China\\
$^2$ College of Physics and Electronic Information, Qinghai Normal University, Xining, People's Republic of China\\
$^3$ Department of  Physics and Institute of Theoretical Physics, Nanjing Normal University, Nanjing 210097, People's Republic of China
}
\date{\today}

\begin{abstract}
Inspired by the recent observations of the vector charmonium-like states at BES III Collaboration and $\psi(3842)$ at LHCb Collaboration, we comb the $D$ wave charmonium state in the present work. We first evaluate the possibility of $Y(4320)$ as $\psi(3^3D_1)$ by investigating its open charm decays in quark-pair creation model and we find the width of $Y(4320)$ can be reproduced in a reasonable parameter range. Moreover, we take $\psi(3770)$, $\psi(4160)$ and $Y(4320)$ as the scale of $1D$, $2D$ and $3D$ charmonia to estimate the open charm decays of other $D$ wave charmonia. The total and partial widths of  $D$ wave charmonium states have been predicted, which could be  tested by further measurements at LHCb and Belle II Collaborations.
\end{abstract}
\pacs{14.40.Pq, 13.20.Gd, 12.39.Fe}

\maketitle

\section{introduction}
By analyzing the precise cross sections for $e^+ e^- \to \omega \chi_{c0} $\cite{Ablikim:2014qwy}, $e^+ e^- \to \pi^+ \pi^- J/\psi $ \cite{Ablikim:2016qzw}, $e^+ e^- \to \pi^+ \pi^- h_c$ \cite{BESIII:2016adj}  and $e^+ e^- \to \pi^+  D^0 D^{\ast-} $ \cite{open-bes}, the BESIII Collaboration reported a series of vector charmonium-like states, which are $Y(4220)$, $Y(4320)$ and $Y(4390)$. The charmonium-like state $Y(4220)$ have been reported in $\chi_{c0} \omega$, $\pi^+ D^0 D^{\ast -}$, $\pi^+ \pi^- h_c$ and $\pi^+\pi^- J/\psi$ channels at present. Its width were reported to be around 40 MeV by analyzing the cross sections for $e^+ e^- \to \chi_{c0} \omega$ and $e^+ e^- \to \pi^+ \pi^- J/\psi$, while it were measured to be about 70 MeV in the cross sections for $e^+ e^- \to \pi^+ \pi^- h_c$  process. As for $Y(4320)$, it was a broad charmonium-like state and only reported in $\pi^+ \pi^- J/\psi$ process.  The charmonium-like state $Y(4390)$ is also a broad state and was observed in the spin flipped $\pi^+ \pi^- h_c$ channel.

These newly observed charmonium-like states make resonances with $J^{PC}=1^{--}$ between $4.0\sim 4.5 $ GeV overcrowed and the nature of these charmonium-like states becomes  an intriguing question. As for $Y(4220)$, it has been observed in various channels. In the $\pi^+ \pi^- J/\psi$ channel, a structure, $Y(4260)$ was firstly reported by BaBar Collaboration \cite{Aubert:2005rm} and then confirmed by Belle Collaboration \cite{Yuan:2007sj}. Recent precise analysis from BESIII Collaboration indicates the structure $Y(4260)$ should contain two charmonium-like state, $Y(4220)$ and $Y(4320)$ \cite{Ablikim:2016qzw}. The former one is consistent with the one observed in the channels of $\chi_{c0} \omega$, $\pi^+ \pi^- h_c$ and $\pi^+ D^0 D^{\ast -}$.  Since $Y(4260)/Y(4220)$ is close to $D_1(2420) \bar{D}$ threshold, it could be considered as a molecular state composed of $D_1(2420) \bar{D}$ \footnote{The charge conjugate states are implied throughout this work}\cite{Ding:2008gr, Cleven:2016qbn, Chen:2016byt, Xue:2017xpu, Cleven:2013mka, Dong:2013kta}.  While, the QCD sum rule estimations indicate that $Y(4260)$ could be a mixed charmonium-tetraquark state \cite{Dias:2012ek, Wang:2016mmg}.

Before the observations of $Y(4220)$, we  predicted a narrow $\psi(4S)$ around 4.2 GeV in Ref.  \cite{He:2014xna}, while $\psi(4415)$ was considered as $\psi(5S)$. After the observation of $Y(4220)$ in the $\chi_{c0} \omega$ channel, The possibility of $Y(4220)$ as $\psi(4S)$ was further evaluated  \cite{Chen:2014sra, Chen:2015bma}. As for $Y(4390)$, it is only observed in the $\pi^+ \pi^- h_c$ channel. In Refs. \cite{Chen:2017abq, He:2017mbh},  the possibility of $Y(4390)$ as a $D^\ast D_1(2420)$ molecular state were investigated. While in Ref. \cite{Chen:2017uof}, the lineshapes of the cross sections for $e^+ e^- \to \pi^+ \pi^- J/\psi,\ \pi^+ \pi^- h_c,\ \pi^+ D^- D^{\ast -}$ could be well reproduced by interferences of the well established charmonia $\psi(4160)$ and $\psi(4415)$ as well as $Y(4220)$.

As for $Y(4320)$, it was also observed in the $\pi^+ \pi^- J/\psi$ channel firstly. Actually, in the $\pi^+ \pi^- \psi(2S)$ channel, there exists a charmonium-like state $Y(4360)$ near the newly observed $Y(4320)$ \cite{Aubert:2007zz, Wang:2007ea}.  The mass  of $Y(4360)$ was fitted to be $4324 \pm 24$ MeV by BaBar Collaboration  \cite{Aubert:2007zz}, which is consistent with the mass of $Y(4320)$. In addition, with recent precise data, the analysis in Ref. \cite{Zhang:2017eta} also indicates that the charmonium-like states $Y(4360)$ in the $\pi^+ \pi^- \psi(2S)$  channel and $Y(4320)$ in the $\pi^+ \pi^- J/\psi$ channel should be the same state. 

\begin{table*}
\caption{Mass spectra and $R$ values of $D$-wave charmonia. SP, GI and MGI refer to the screen potential model \cite{Li:2009zu}, Godfrey-Isgur relativistic quark model \cite{Godfrey:1985xj} and modified Godfrey-Isgur relativistic quark model \cite{Wang:2019mhs}. The values in the bracket are the effective $R$ values of the corresponding states in unit of $\rm GeV^{-1}$. \label{Tab:Dwave}}
\begin{tabular}{p{2.5cm}<{\centering} p{3.5cm}<{\centering} p{3.cm}<{\centering} p{3cm}<{\centering} p{3cm}<{\centering}}
\toprule[1pt]
States & Experiment &  SP Model \cite{Li:2009zu}  & GI  Model \cite{Godfrey:1985xj}  & MGI Model \cite{Wang:2019mhs}\\
$\eta_{c2}(1D)$  &---&3796 &  3837 & 3848\\ 
$\psi_1(1D)$ & $3773.13 \pm 0.15$  \cite{Tanabashi:2018oca} & 3783 (2.59)  & 3821 (1.84) & 3830 (1.88)   \\
$\psi_2(1D)$ & $3822.2 \pm 1.2$ \cite{Tanabashi:2018oca} & 3798 & 3838 & 3848 \\
$\psi_3(1D)$  & $3842.71 \pm 1.6 \pm0.12 $ \cite{LHCbnew} & 3799  & 3846 & 3858 \\
\midrule[1pt]
$\eta_{c2}(2D)$  &--- & 4099 & 4207 & 4137\\ 
$\psi_1(2D)$ & $4191\pm 5$ \cite{Tanabashi:2018oca} & 4089 (3.12) & 4197 (2.09) & 4125 (2.38) \\
$\psi_2(2D)$ &---& 4100  & 4209 & 4137  \\
$\psi_3(2D)$  &---& 4103 & 4215 & 4144\\
\midrule[1pt]
$\eta_{c2}(3D)$ &---&4326  & 4531 & 4343\\ 
$\psi_1(3D)$ & --- &4317 (3.59) & 4522 (2.24) & 4334 (2.85)\\
$\psi_2(3D)$   &---& 4327 & 4532 & 4343\\
$\psi_3(3D)$  &---& 4331 & 4536 & 4348\\
\bottomrule[1pt]
\end{tabular}

\end{table*}

In our previous work, we have categorized $Y(4220)$ as $\psi(4S)$  and $\psi(4415)$ as $\psi(5S)$  \cite{He:2014xna, Chen:2014sra, Chen:2015bma}.  In such a scenario, there are no additional room left for $Y(4320)$  in the $S$-wave vector charmonium and in the vicinity of $Y(4320)$, there is no charmed mesons pair threshold. However, if one further checks the charmonium spectroscopy, one can find that in the $D$-wave charmonium sector, $\psi(3770)$ and $\psi(4160)$ are well established as $\psi(1^3D_1)$ and $\psi(2^3D_1)$ states, respectively. The higher $D$-wave vector charmonia have not been observed experimentally.  On the theoretical side, the masses of $D$ wave charmonia have been predicted in the quark model as shown in Table \ref{Tab:Dwave}. One can find the mass of  $\psi(3^3D_1)$ was predicted to be $4519$ MeV by the relativistic quark model \cite{Godfrey:1985xj}. However, for the higher charmonia, the couple channel effects will shift their mass  to the open-charm threshold \cite{Li:2009ad, Kalashnikova:2005ui,  Ortega:2009hj}, thus the predicted mass of $\psi(3^3D_1)$ in Ref. \cite{Godfrey:1985xj} should be too large since the coupled-channel effects are not included.  In Refs.  \cite{Li:2009zu, Wang:2019mhs}, the screened potential model were employed to depict the couple channel effect in the charmonium, the predicted mass of  $\psi(3^3D_1)$ is $4317$ MeV and $4334$ MeV, respectively, which is well consistent with the one of $Y(4320)$. Thus, $Y(4320)$ could be a good candidate of $\psi(3^3D_1)$ state.

Moreover, very recently, the LHCb collaboration reported their measurements of the near threshold $D\bar{D}$ spectroscopy \cite{LHCbnew}. In the $D\bar{D}$ mass spectrum, the $D$ wave charmonium $\psi(3770)$ was observed in the hadronproduction process for the first time \cite{LHCbnew}. In the same spectroscopy, a new narrow state (named $\psi(3842)$ hereafter) was reported.
As shown in Table. \ref{Tab:Dwave}, the mass of this newly observed state is consistent with one of $\psi(1^3D_3)$ state predicted by quark model \cite{Godfrey:1985xj, Li:2009zu, Wang:2019mhs} and the narrow width could result from the higher partial wave suppression since $\psi(1^3D_3)$ decays into $D\bar{D}$ via a $F$ wave with $L=3$.  Moreover, another $D$-wave charmonia candidate, $\psi(3823)$, was firstly observed by Belle Collaboration \cite{Bhardwaj:2013rmw} and then confirmed by BES III Collaboration \cite{Ablikim:2015dlj}. Considering $\psi(3823)$ as $\psi_2(1D)$ state, together with the newly observed $\psi(3842)$ as $\psi_3(1D)$ state, the $D$ wave ground spin triplets have been well established. As for $2D$ charmonia, one can find only $\psi_1(2D)$ state has been observed experimentally. Thus, searching the missing highly excited $D$ wave charmonia experimentally will be intriguing. Unlike to the electron-positron annihilation process, the states produced in the hadronproduction process have more possibility of $J^{PC}$ quantum numbers, while states involved in the electron-positron annihilation process have fixed $J^{PC}$ quantum numbers, which are $1^{--}$. Thus, the hadronproduction process in the LHCb Collaboration provide us a powerful platform of searching for charmonium states with various $J^{PC} $ quantum numbers, which includes the missing highly excited $D$ wave charmonia.

On the theoretical side, it will be intriguing to comb the $D$ wave charmonium states. In the present work, we take $\psi(3770)$, $\psi(4160)$ and $Y(4320)$ as the  $\psi(1^3D_1)$, $\psi(2^3D_1)$ and $\psi(3^3D_1)$ charmonia, and take these states as scales to investigate the open charm decays of other $D$ wave charmonium states, which could, to some extend, cancel the uncertainties of quark model.

This work is organized as follows. After introduction,  a short review of quark pair creation model and the formula of open-charm decays of $D$ wave charmonium states are presented in Section II. Our numerical results  and discussions are given in Section III. Section IV is devoted to summary.

\section{Quark pair Creation model and open charm decays of $\psi(^3D_1)$ charmonium}

\subsection{Review of quark pair creation model}
Here, we adopt the quark pair creation (QPC) model (also named $^3P_0$ model since the $J^{PC}$ quantum numbers of the quark pair created from the vacuum are $0^{++}$) to estimate the open charm decays of charmonia. The QPC model was first proposed by Micu \cite{Micu:1968mk, LeYaouanc:1972vsx, LeYaouanc:1973ldf, LeYaouanc:1977fsz}  and then widely used to estimate the OZI allowed strong decay processes \cite{Liu:2009fe, Close:2005se, Song:2014mha, Godfrey:2015dia, Chen:2016iua, Barnes:2003vb, Wang:2019mhs, Song:2015fha, Song:2015nia}. In the QPC model, the related $S-$ matrix of $A\to BC$ process reads,
\begin{eqnarray}
\langle BC \left| S \right| A \rangle = I- i 2\pi \delta(E_f-E_i)  \langle BC \left|\mathcal{T}\right| A\rangle,   \label{Eq:SMat}
\end{eqnarray}
where the transition operator $\mathcal{T}$ is,
\begin{eqnarray}
\mathcal{T} =&&-3 \gamma \sum_m \langle 1m;1-m|00\rangle \int d \k_3 \k_4 \delta^3 (\k_3 +\k_4) \nonumber \\
&&\times \mathcal{Y}_{1m}\left(  \frac{\k_3-\k_4}{2} \right) \chi_{1,-m}^{34} \varphi_0^{34} \omega_{0}^{34} d_{3i}^{\dagger} (\k_3) b_{4j}^{\dagger} (\k_4), \label{Eq:Tran}
\end{eqnarray}
where $\mathcal{Y}_{1m}(\k )=|\k| Y_{1m}(\theta ,\phi)$, $\chi_{1,-m}^{34}$,  $\varphi_0^{34}= (u\bar{u}+d\bar{d}+s\bar{s})/\sqrt{3}$ and $\omega_0^{34} =\delta_{\alpha_3 \alpha_4}$ are the space, spin, flavor and color parts of the wave functions, respectively. $\alpha_3$ and $\alpha_4$ are the color indexes of the created quark pair. In the QPC model, the parameter $\gamma $ is introduced to represent the strength of the quark-antiquark pair creation from the vacuum and it could be fixed by fitting the decay data.  In the present work, we take $\gamma=6.3$ for the up/down quark pair and $\gamma_s =\gamma/\sqrt{3}$ for strange quark pair creation \cite{Chen:2016iua, Liu:2009fe}.

In the initial rest frame, the matrix element of the transition operator is
\begin{eqnarray}
&&\langle BC \left| \mathcal{T} \right| A\rangle  = \sqrt{8 E_A E_B E_C} \gamma \sum_{\substack{M_{L_A}, M_{L_B}, M_{L_C}, \\M_{S_A}, M_{S_B}, M_{S_C}}}  \langle 1m, 1-m|00\rangle  \nonumber\\
&&\hspace{5mm}\times \langle L_A, M_{L_A}, S_A M_{S_A}|J_A, M_A\rangle
\langle L_B, M_{L_B}, S_B M_{S_B}|J_B, M_B\rangle  \nonumber\\
&&\hspace{5mm}\times \langle L_C, M_{L_C}, S_C M_{S_C} |J_C, M_C\rangle  \langle \varphi_{B}^{13} \varphi_{C}^{24} | \varphi_{A}^{12} \varphi_0^{34} \rangle \nonumber\\
&&\hspace{5mm}\times \langle \chi_{S_B M_{S_B}}^{13} \chi_{S_C M_{S_C}}^{24} | \chi_{S_A M_{S_A}}^{12} \chi^{34}_{1-m} \rangle I_{M_{L_B} M_{L_C}}^{M_{L_A} m} (\mathbf{K}),
\end{eqnarray}
where $\langle \varphi_{B}^{13} \varphi_{C}^{24} | \varphi_{A}^{12} \varphi_0^{34} \rangle$ and $\langle \chi_{S_B M_{S_B}}^{13} \chi_{S_C M_{S_C}}^{24} | \chi_{S_A M_{S_A}}^{12} \chi_{1-m}^{34}\rangle$ are the flavor matrix element and spin matrix element, respectively. While the color matrix element $\langle \omega_{B}^{13} \omega_{C}^{24} | \omega_{A}^{12} \omega_0^{34} \rangle =1/3$ cancels out the factor $3$ in the transition operator defined in Eq. (\ref{Eq:Tran}).  The matrix element of the spatial part reads
\begin{eqnarray}
&& I^{M_{L_A},m}_{M_{L_B},M_{L_C}}(\textbf{K}) = \int\!\rm
d\mathbf{k}_1\rm d\mathbf{k}_2\rm d\mathbf{k}_3\rm
d\mathbf{k}_4\,\delta^3(\mathbf{k}_1+\mathbf{k}_2)\delta^3(\mathbf{k}_3+\mathbf{k}_4) \nonumber\\ &&
\hspace{5mm} \times \delta^3
(\textbf{K}_B-\mathbf{k}_1-\mathbf{k}_3)\delta^3(\textbf{K}_C-\mathbf{k}_2-\mathbf{k}_4) \Psi^*_{n_B L_B
M_{L_B}}(\mathbf{k}_1,\mathbf{k}_3)\nonumber\\
&&\hspace{5mm}  \times\Psi^*_{n_C L_C
M_{L_C}}(\mathbf{k}_2,\mathbf{k}_4)
\Psi_{n_A L_A M_{L_A}}(\mathbf{k}_1,\mathbf{k}_2)
\mathcal{Y}_{1m}\Big(\frac{\mathbf{k}_3-\mathbf{k}_4}{2}\Big),
\label{Eq:I}
\end{eqnarray}
which reflects the overlap of the  spatial wave functions of the initial state and final states. The amplitude of the decay process is
\begin{eqnarray}
\langle BC \left| \mathcal{T} \right| A\rangle  =\delta^3(\mathbf{K}_B+\mathbf{K}_C-\mathbf{K}_A) \mathcal{M}^{M_{J_A} M_{J_B} M_{J_C}}.
\end{eqnarray}
By the Jacobi-Wick rotation, the amplitude can be transformed into partial wave amplitude, which is,
\begin{eqnarray}
\mathcal{M}^{JL}(A\to BC) &=&\frac{\sqrt{2L+1}}{2J_A +1} \sum_{M_{J_B},M_{J_C}} \langle L0 JM_{J_A} | J_A M_{J_A}\rangle \nonumber\\ &&\times  \langle J_B M_{J_B } J_C M_{J_C} | J_A M_{J_A}  \rangle \mathcal{M}^{M_{J_A} M_{J_B} M_{J_C}}. \label{Eq:PWA}\nonumber\\
\end{eqnarray}
In terms of the partial wave amplitude, the partial width is
\begin{eqnarray}
\Gamma = \pi^2 \frac{\left|\mathbf{K} \right|}{m_{A}^2} \sum_{JL} \left| \mathcal{M}^{JL}\right| ,
\end{eqnarray}
where $\left| \mathbf{K}\right| = \lambda^{1/2} (m_A^2, m_B^2, m_C^2)$ with the K$\mathrm{\ddot{a}}$llen function $\lambda(x,y,z)=x^2 +y^2 +z^2 -2xy -2yz -2xz$.

\subsection{Open charm decays of $D$-wave charmonia}

\begin{table}[htb]
\centering
\caption{The masses and $R$ values of the involved mesons. Here $(\pm)$ and $(0)$ indicate the charge of the mesons.  \label{Tab:mass}}
\begin{tabular}{p{1.5cm}p{3.5cm}<{\centering}p{2.5cm}<{\centering}}
\toprule[1pt]
Meson & Mass (MeV)  &  $R\ (\mathrm{GeV}^{-1})$ \cite{Godfrey:1986wj} \\
\midrule[1pt]
$D$  &       $1864.83 (0), 1869.58(\pm)$  & $1.52$\\
$D^\ast$  &  $2006.85(0), 2010.26(\pm)$  & $1.85$\\
$D_0(2400)$      &  $2318(0) 2351(\pm)$ & $1.85$\\
$D_1(2420)$    & $2420.8(0),  2423.2 (\pm)$ &$2.00$\\
$D_1^\prime(2430)$  & $2427 (0),\ 2427(\pm) $  & $2.00$ \\
$D_2(2460)$  & $2460.7 (0),\ 2465.4(\pm) $  & $2.00$ \\
$D_s$        & $1968.28 (\pm)$ & $1.41$ \\
$D_s^\ast$ &$2112.1 (\pm)$& $1.69$\\
\midrule[1pt]
$\psi(3770)$ &$3773.13$ &---\\
$\psi_3(3842)$& $3842.71$&---\\
$\eta_{c2}(2D)$ & 4201& ---\\
$\psi(4160)$ &$4191$ &---\\
$\psi_2(2D)$ & $4203$ &---\\
$\psi_3(2D)$ & $4209$ &---\\
$\eta_{c2}(3D)$ & $4330$ & ---\\
$Y(4320)$    &$4320.0$&---\\
$\psi_2(3D)$ & $4330$ &---\\
$\psi_3(3D)$ & $4335$ &---\\
\bottomrule[1pt]
\end{tabular}
\end{table}

\begin{table*}[t]
\caption{The open charm decay modes of  $D$ wave charmonia. \label{Tab:mode} }
\begin{tabular}{ccccccccccccc}
\toprule[1pt]
  Channel   &$\eta_ {c2}(1D)$ & $\psi_1(1D)$ & $\psi_2(1D)$ & $\psi_3(1D)$ & $\eta_{c2 }(2D)$ & $\psi_1(2D)$ & $\psi_2(2D)$ & $\psi_3(2D)$ & $\eta_ {c2}(3D)$ & $\psi_1(3D)$ & $\psi_2(3D)$ & $\psi_3(3D)$  \\
 \midrule[1pt]
   $ D \bar{D} $ &  $\dots $ &  $\checkmark $ &  $\dots $ &  $\checkmark $ &  $\dots $ &  $\checkmark $ &  $\dots $ &  $\checkmark $ &  $\dots $ &  $\checkmark $ &  $\dots $ &  $\checkmark $  \\
   $ D \bar{D}^{\ast} $ &  $\dots $ &  $\dots $ &  $\dots $ &  $\dots $ &  $\checkmark $ &  $\checkmark $ &  $\checkmark $ &  $\checkmark $ &  $\checkmark $ &  $\checkmark $ &  $\checkmark $ &  $\checkmark $  \\
   $ D^{\ast} \bar{D}^{\ast} $ &  $\dots $ &  $\dots $ &  $\dots $ &  $\dots $ &  $\checkmark $ &  $\checkmark $ &  $\checkmark $ &  $\checkmark $ &  $\checkmark $ &  $\checkmark $ &  $\checkmark $ &  $\checkmark $  \\
   $D_s^+ D_s^- $ &  $\dots $ &  $\dots $ &  $\dots $ &  $\dots $ &  $\dots $ &  $\checkmark $ &  $\dots $ &  $\checkmark $ &  $\dots $ &  $\checkmark $ &  $\dots $ &  $\checkmark $  \\
   $D_s^+ D_s^{\ast-} $ &  $\dots $ &  $\dots $ &  $\dots $ &  $\dots $ &  $\checkmark $ &  $\checkmark $ &  $\checkmark $ &  $\checkmark $ &  $\checkmark $ &  $\checkmark $ &  $\checkmark $ &  $\checkmark $  \\
   $D_s^{\ast +}D_s^{\ast-} $ &  $\dots $ &  $\dots $ &  $\dots $ &  $\dots $ &  $\dots $ &  $\dots $ &  $\dots $ &  $\dots $ &  $\checkmark $ &  $\checkmark $ &  $\checkmark $ &  $\checkmark $  \\
   $ D\bar{D}_0 $ &  $\dots $ &  $\dots $ &  $\dots $ &  $\dots $ &  $\checkmark $ &  $\dots $ &  $\checkmark $ &  $\dots $ &  $\checkmark $ &  $\dots $ &  $\checkmark $ &  $\dots $  \\
   $ D\bar{D}_1(2420) $ &  $\dots $ &  $\dots $ &  $\dots $ &  $\dots $ &  $\dots $ &  $\dots $ &  $\dots $ &  $\dots $ &  $\checkmark $ &  $\checkmark $ &  $\checkmark $ &  $\checkmark $  \\
   $ D\bar{D}_1(2430) $ &  $\dots $ &  $\dots $ &  $\dots $ &  $\dots $ &  $\dots $ &  $\dots $ &  $\dots $ &  $\dots $ &  $\checkmark $ &  $\checkmark $ &  $\checkmark $ &  $\checkmark $  \\
   $ D^{\ast}\bar{D}_0(2400) $ &  $\dots $ &  $\dots $ &  $\dots $ &  $\dots $ &  $\dots $ &  $\dots $ &  $\dots $ &  $\dots $ &  $\checkmark $ &  $\dots $ &  $\checkmark $ &  $\checkmark $  \\
   $D_s^+D_{s0}^-(2317) $ &  $\dots $ &  $\dots $ &  $\dots $ &  $\dots $ &  $\dots $ &  $\dots $ &  $\dots $ &  $\dots $ &  $\checkmark $ &  $\dots $ &  $\checkmark $ &  $\dots $ \\
\bottomrule[1pt]
\end{tabular}
\end{table*}

In the present work, we perform a system estimation of the open charm decays of $D$ wave charmonia. To evasion the uncertainties of quark model, we take $\psi(3770)$, $\psi(4160)$ and $Y(4320)$ as the scale of $1D$, $2D$ and $3D$ charmonium states.  The masses of the involved charmonium states and charmed mesons are listed in Table \ref{Tab:mass}. As for the charmed mesons and already established charmonia, i.e., $\psi(3770)$ and $\psi(4160)$,  we adopt the center values of the PDG average \cite{Tanabashi:2018oca}.  As for $Y(4320)$ and $\psi_3(3842)$, we take the measurement one in Ref. \cite{Ablikim:2016qzw, LHCbnew}.  It's interesting to notice the mass splitting between the same spin multiplets are predicted to be very similar for different quark model \cite{Li:2009zu, Wang:2019mhs, Godfrey:1985xj}. 
For example, the mass splitting $\Delta m_{a} = m_{\eta_{c2}(2D)}- m_{\psi_1(2D)} $ are predicted to 10, 12 and 10 MeV for GI model, MGI model and SP model, respectively. By using the mass splitting estimated in SP model and taking $\psi (4160)$ as the scale of $2D$ states, the masses of the missing $2D$ states can be estimated, for example, $m_{\eta_{c2}(2D)} =m_{\psi(4160)}+\Delta m_{a} =4201 \ \rm{MeV}$.  In the same way, the masses of the missing $3D$ states can be evaluated by taking $Y(4320)$ as the scale.

 Considering the $J^{PC}$ conservation and kinetics limit, we list all the possible open charm decay modes of $D$ wave charmonia in Table \ref{Tab:mode}. As for ground states, $\psi_1(1D)$ and $\psi_3(1D)$ can decay into $D\bar{D}$ via $P$ wave and $F$ wave, respectively, while $\eta_{c2}(1D)$ and $\psi_2(1D)$ have no open charm decay mode, although they are above the threshold of $D\bar{D}$. With Eqs. (\ref{Eq:SMat})-(\ref{Eq:PWA}), one can get the partial wave amplitudes of the involved process as shown in Table \ref{Tab:mode}. The estimated particular expressions of these partial wave amplitudes are listed in Table \ref{Tab:amp1}-\ref{Tab:amp4} in Appendix \ref{Sec:App}.

For $3D$ charmonia, their masses are above the threshold of $D_1^\prime(2430) \bar{D}$ and $D_1(2420) \bar{D}$. The charmed meson $D_1^\prime(2430)$ and $D_1(2420)$  are the mixture of the $1^3P_1$ and $1^1P_1$ states and the mixing scheme is,
\begin{eqnarray}
\left(
\begin{array}{c}
|D_1^\prime(2430) \rangle \\
|D_1(2420) \rangle
\end{array}
\right)=
\left(
\begin{array}{cc}
\cos \theta & \sin \theta \\
-\sin \theta & \cos \theta
\end{array}
\right)
\left(
\begin{array}{c}
|1^1P_1 \rangle \\
|1^3P_1 \rangle
\end{array}
\right),
\end{eqnarray}
where the mixing angle $\theta =-54.7^\circ$, which is determined by the heavy quark limit \cite{Godfrey:1986wj, Barnes:2002mu, Matsuki:2010zy}.

\section{Numerical Results and discussions}

With above preparations, we could investigate the open charm decays of the $D$ wave charmonia. In Eq. (\ref{Eq:I}), the spatial wave functions of the mesons are involved. In principle, these wave functions could be estimated by the constitute quark model. However, as we discussed in the introduction, there exist some uncertainties in the quark models. Thus,  in the present work, we employ the simple harmonic oscillator wave function to simulate the spatial distribution of the quark-antiquark in meson. the detailed form of the spatial wave function in the momentum representation is
\begin{eqnarray}
\Psi_{n\ell m_\ell} (R, \k)&=&  \frac{(-1)^n (-i)^\ell R^{3/2}}{\sqrt[4]{\pi}}\sqrt{\frac{2^{\ell -n+2} (2\ell +2n +1)!!}{n! (2\ell +1)!!^2}} (k R)^\ell  \nonumber\\ &&\times F\left(-n,\ell+\frac{3}{2}, R^2 k^2\right) e^{-R^2 k^2 /2} Y_{\ell ,m_\ell}(\hat{\k})
\end{eqnarray}
where $n,\ \ell$ and $m_\ell$ are the radial, angular momentum and magnetic quantum numbers, respectively. $F(-n, \nu, x)$ and $Y_{\ell m_{\ell}}$ indicate the hypergeometric function and spherical harmonic function, respectively.

In the spatial wave function, a parameter $R$ is introduced. As for the lowest charmed mesons, the predictions of the relativistic quark model are well consistent with the experimental measurements. Thus, in the present work,  the values of parameter $R$ for the charmed and charmed-strange mesons are fixed such that it reproduces the root mean square radius estimated by the relativistic quark model \cite{Godfrey:1986wj}. In Ref. \cite{Close:2005se, Song:2014mha, Godfrey:2015dia, Barnes:2003vb, Chen:2016iua}, the simple harmonic oscillator wave function with a parameter $R$ has been used to investigate the decay behavior of mesons and the estimated results could well reproduce the corresponding experimental data, which proves such an approach is reliable to investigate the strong decays of the hadrons.  As for the charmonia, the $R$ values are  quite different in different quark model as shown in Table \ref{Tab:Dwave}. For example, the $R$ value of $1D$ states are estimated to be $2.59$, $1.84$ and $1.88 \ \rm{GeV}^{-1}$ by SP, GI and MGI model, respectively.  It should be noticed that the mass of charm quark is taken as $1.4045 $ GeV and $1.65$ GeV in the SP and MGI models, respectively. The mass spectra and the wave function depend on both the quark mass and the potential between quark and anti-quark. Thus, the large discrepancy of $R$ values in SP and MGI model could  be understood. In the $^3P_0$ model, the constituent quark masses for the charm, up/down, strange quarks are adopted to be 1.60, 0.22 and 0.419 GeV, respectively \cite{Chen:2016iua, Liu:2009fe}. The mass of the charm quark in $^3P_0$ model is very close to the one in MGI model. Thus in the present work, we vary the $R$ values of the charmonia around the one of MGI model to check the $R$ dependence of the decay widths. In addition, similar to the case of determining the masses of the missing $D$ wave states, the $R$ values of $1D$, $2D$ and $3D$ states could be determined such that it could reproduce the  widths of $\psi(3770)$, $\psi(4160)$ and $Y(4320)$, respectively, which could also reduce the uncertainties of quark model.  
The masses and $R$ values of the involved mesons are presented in Table \ref{Tab:mass}. 

\begin{figure}[htb]
\vspace{0.25cm}
\scalebox{1.0}{\includegraphics{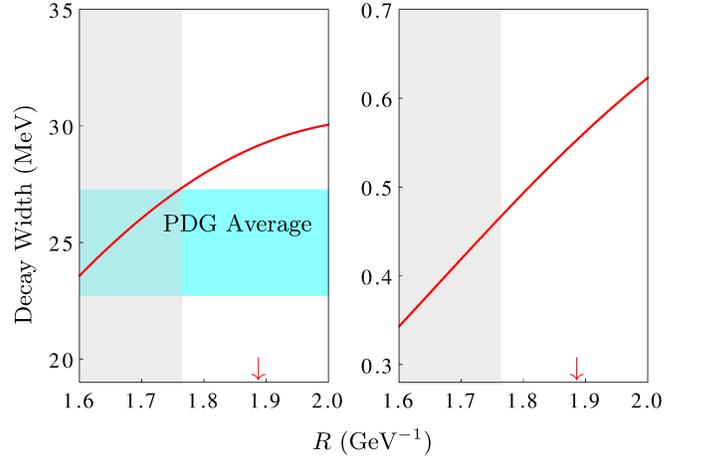}}
\caption{(Color online). The partial width of $\psi(1D) \to D\bar{D}$ (left panel) and $\psi_3(1D) \to D\bar{D}$ (right panel). The cyan band indicates the PDG average of the corresponding partial width. The $R$ value estimated by MGI model are marked by red arrow.  \label{Fig:psi1D}}
\end{figure}

\subsection{Open Charm Decays of $1D$ States}

As for $1D$ charmonia, their masses are all above the $D\bar{D}$ threshold, however, $\eta_{c2}$ and $\psi_2(1D)$ can not decay into $D\bar{D}$ due to $J^P$ quantum numbers violation. As for the $\psi(3770)$, the only open charm decay mode is $D\bar{D}$ due to kinematics limit. The $R$ dependence of the partial width of $\psi(3770)\to D\bar{D}$ is presented in the left panel of Fig. \ref{Fig:psi1D}. The $R$ value estimated by MGI model is marked by the red arrow and with this $R$ value, the partial width of $\psi(3770) \to D\bar{D}$ is evaluated to be 29.1 MeV. The PDG average of the the branching ratio for $\psi(3770) \to D\bar{D} $ is $(93^{+8}_{-9}) \%$ and the width of the $\psi(3770)$ is $27.2\pm 1$ MeV \cite{Patrignani:2016xqp}. Thus, the measured partial width of $\psi(3770) \to D\bar{D}$ is $22.7 \sim 27.3 $ MeV, which indicates the partial width with $R$ value in MGI model is approximately consistent with the experimental measurement. Moreover,  we vary $R$ value from $1.6\ \mathrm{GeV}^{-1}$ to $2.0\ \mathrm{GeV}^{-1}$ and  find that the estimated partial width of $\psi(3770) \to D \bar{D}$ with $R=1.6 \sim 1.76 \ \mathrm{GeV}^{-1}$ could well reproduce the experimental measurement. Taking $\psi(3770)$ as a scale of $1D$ charmonia, the partial width of $\psi_3(1D) \to D\bar{D}$ is $0.34 \sim 0.46$ MeV, which is consistent with the theoretical estimations in Ref. \cite{Barnes:2005pb, Deng:2016stx} and safely under the measured width from the LHCb Collaboration \cite{LHCbnew}

\begin{figure}[htb]
\vspace{0.25cm}
\scalebox{0.8}{\includegraphics{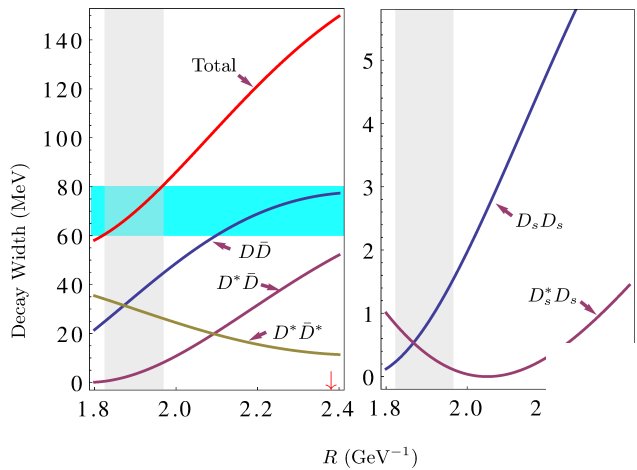}}
\caption{(Color online). Partial and total widths of $\psi(4160)$. The cyan horizontal band is the PDG average of the total width and the light vertical grey band is the $R$ range determined by the overlap of theoretical estimation and experimental data. The $R$ value estimated by MGI model are marked by red arrow. \label{Fig:psi12d}}
\end{figure}

\subsection{Open Charm Decays of $2D$ States}

The $R$ dependent partial and total widths of $\psi(4160)$ are presented in Fig. \ref{Fig:psi12d}.  By taking the $R$ value determined in MGI model, the width of $\psi(4160)$ is estimated to be 147.4 MeV, which is about two time larger than the PDG average one, i.e., $70 \pm 10$ MeV. It should noticed that the mass of $\psi(2D)$ is estimated to be 4.125 GeV in the MGI model, which is much smaller than the measured one, thus we discuss the the decay behavior of $2D$ states in the $R$ range determined by comparing the estimated width with the experimental data \cite{Patrignani:2016xqp}, which is  $R= (1.82 \sim 1.97) \ \mathrm{GeV}^{-1}$. Moreover, the determined $R$ value of $\psi(4160)$ is a bit larger than the one of $\psi(3770)$, which is consistent with the expectation.  In this $R$ range, our results indicates the $\psi(4160)$ dominantly decays into $D\bar{D}$, $D^\ast \bar{D}$ and $D^\ast \bar{D}^\ast$, while the partial widths of $D_s^+ D_s^-$ and $D_s^+ D_s^{\ast -}$  are less than 1 MeV. In this $R$ range, the ratios of the partial widths of open charmed processes are estimated to be
\begin{eqnarray}
\Gamma(\psi(4160)\to D\bar{D})\over \Gamma(\psi(4160) \to D^\ast \bar{D}^\ast) &=&0.71 \sim 1.72 \nonumber\\
\Gamma(\psi(4160)\to D^\ast \bar{D})\over\Gamma(\psi(4160) \to D^\ast \bar{D}^\ast)&=&0.01 \sim 0.31 \nonumber
\end{eqnarray}
These ratios are evaluated to be 0.46/0.01 and 0.2/0.05 by the QPC model with relativistic quark model  and linear potential model, respectively\cite{Barnes:2005pb, Gui:2018rvv}.  In Ref. \cite{Eichten:2005ga} , by using the Connell coupled- channel mode, the ratios are determined to be 0.08 and 0.16.  On the experimental side, the BaBar collaboration performed a measurement of the exclusive production of $D\bar{D}$, $D^\ast \bar{D}$ and $D^\ast \bar{D}^\ast$, the ratios were measured to be $0.02 \pm 0.03 \pm 0.02$ and $0.34 \pm 0.14 \pm 0.05$ \cite{Aubert:2009aq}, respectively, which is different from the QPC model estimations in the present work. It should be noticed that in Ref. \cite{Aubert:2009aq}, the data are fitted with three charmonia with fixed mass and width, which are $\psi(4040)$, $\psi(4160)$ and $\psi(4415)$. From the current situation, there should exist more vector states in this energy range and thus the fitted results will be changed if more states are included.  Moreover, in the analysis, the mass and width of $\psi(4160)$ are fixed to be $4153$ MeV and $103$ MeV, respectively \cite{Amsler:2008zzb}. The values of the resonance parameters used in Ref. \cite{Aubert:2009aq}  are much different from latest PDG average, which are $4191$ MeV and $70$ MeV, respectively \cite{Patrignani:2016xqp}.  We expect the new precise measurement and analysis of the open charm decays of $\psi(4160)$ at BESIII, BelleII and LHCb could determine these ratios and test the results in the present work.

\begin{figure}[htb]
\vspace{0.25cm}
\scalebox{0.8}{\includegraphics{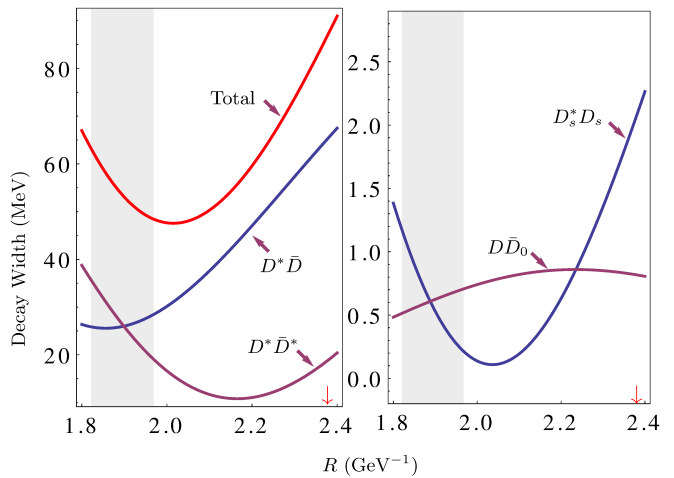}}
\caption{(Color online). Partial and total widths of $\eta_{c2}(2D)$. The light grey band is the $R$ range determined by the comparison of $\psi(4160)$ total width with the experimental data. The $R$ value estimated by MGI model are marked by the arrow. \label{Fig:etac22d}}
\end{figure}

Taking the $R$ range determined by the width of $\psi(4160)$, we can investigate the open charm decays of other $2D$ charmonium states.  As for $\eta_{c2}(2D)$, the partial and total widths depending on $R$ value are presented in Fig. \ref{Fig:etac22d}. The total width of $\eta_{c2}(2D)$ is estimated to be $48\sim 64 $ MeV. The dominant decay modes are $D^\ast \bar{D}^\ast$ and $D^\ast \bar{D}$ and the ratio of these two decay channels is estimated to be
\begin{eqnarray}
\frac{\Gamma(\eta_{c2}(2D) \to D^\ast \bar{D})}{\Gamma(\eta_{c2}(2D) \to D^\ast \bar{D}^\ast)} = 0.7 \sim 1.5,
\end{eqnarray}
which indicates the partial width of $\Gamma(\eta_{c2}(2D) \to D^\ast \bar{D})$ and $\Gamma(\eta_{c2}(2D) \to D^\ast \bar{D}^\ast)$ are very similar. As for the $D_s^{\ast+} D_s^{-}$ and $D \bar{D}_0$ modes, their partial widths are less than 1 MeV.

\begin{figure}[htb]
\vspace{0.25cm}
\scalebox{0.8}{\includegraphics{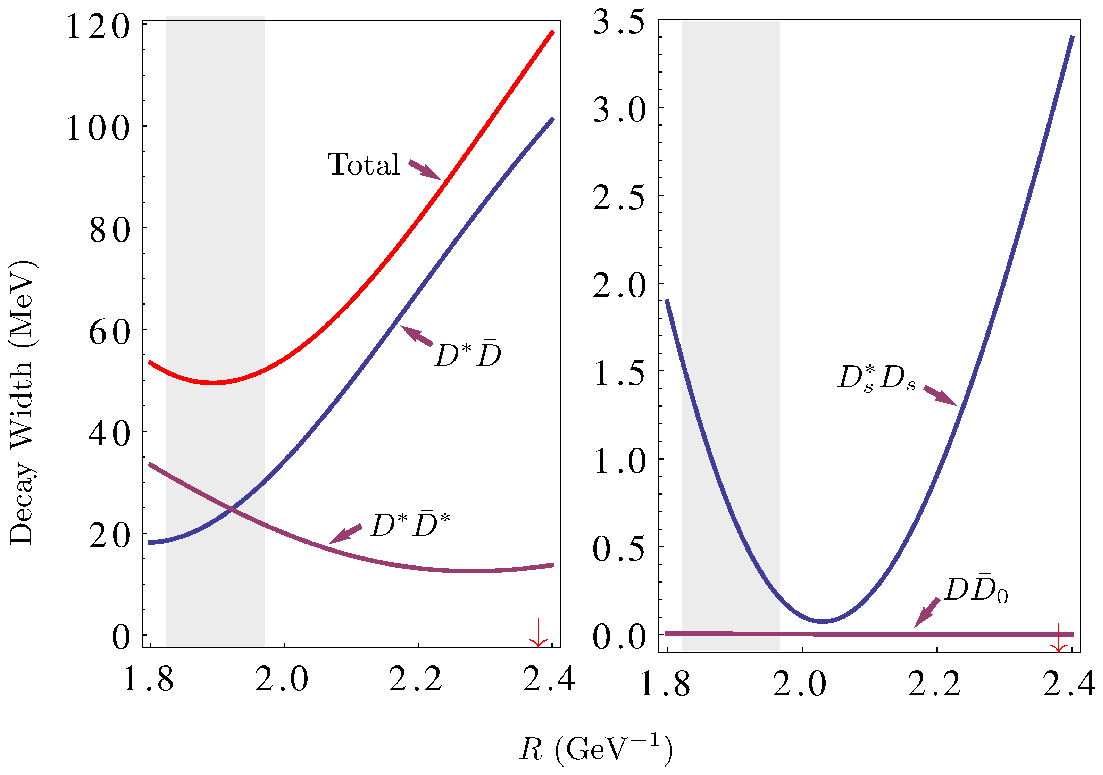}}
\caption{(Color online). The same as Fig. \ref{Fig:etac22d} but for $\psi_2(2D)$ charmonium. \label{Fig:psi22d}}
\end{figure}

\begin{figure}[htb]
\vspace{0.25cm}
\scalebox{0.8}{\includegraphics{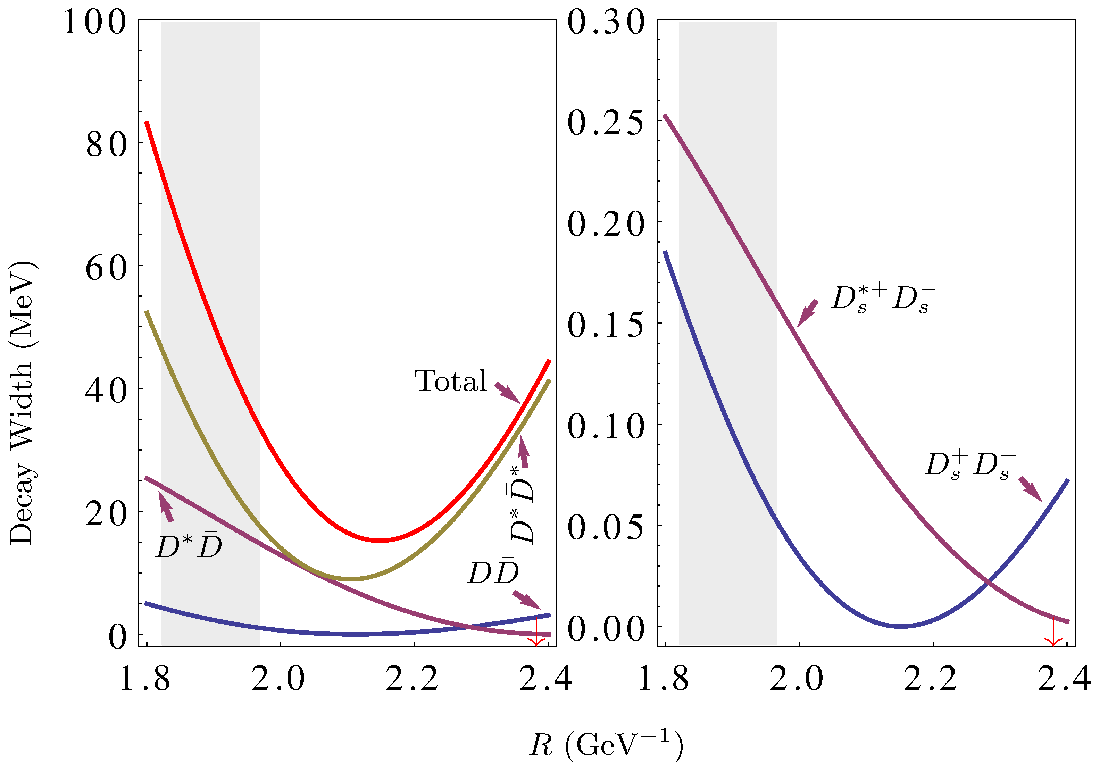}}
\caption{(Color online). The same as Fig. \ref{Fig:etac22d} but for $\psi_3(2D)$ charmonium. \label{Fig:psi32d}}
\end{figure}

As for $\psi_2(2D)$, the total width  is very weakly dependent on the $R$ values in the determined $R$ range,  and it is estimated to be $50 \sim 52$ MeV. Similar to the case of $\eta_{c2}(2D)$, the dominant decay modes of $\psi_2(2D)$ are also $D^\ast \bar{D}$ and $D^\ast \bar{D}^\ast$, and the ratio of the partial widths for these two channels are estimated to be,
\begin{eqnarray}
\frac{\Gamma(\psi_2(2D) \to D^\ast \bar{D})}{\Gamma(\psi_{2}(2D) \to D^\ast \bar{D}^\ast)} = 0.6 \sim 1.4.
\end{eqnarray}
As for $\psi_3(2D)$, the total width are estimated to be $52 \sim 76$ MeV in the determined $R$ range. Such a large width mainly comes from the $D^\ast \bar{D}$ and $D^\ast \bar{D}^\ast$ modes since $\psi_3(2D)$ decays into $D^\ast \bar{D}$ and $D^\ast \bar{D}^\ast$ are also via $P$ wave. The partial widths ratio of these two channel is predicted to be,
\begin{eqnarray}
\frac{\Gamma(\psi_{3}(2D) \to D^\ast \bar{D})}{\Gamma(\psi_{3}(2D) \to D^\ast \bar{D}^\ast)} = 0.5 \sim 0.8.
\end{eqnarray}
Compared with the above two channels, the partial width of $D\bar{D}$ mode is much smaller due to the high partial wave suppression.

\subsection{Open Charm Decays of $3D$ States}

\begin{figure}[htb]
\vspace{0.25cm}
\scalebox{0.8}{\includegraphics{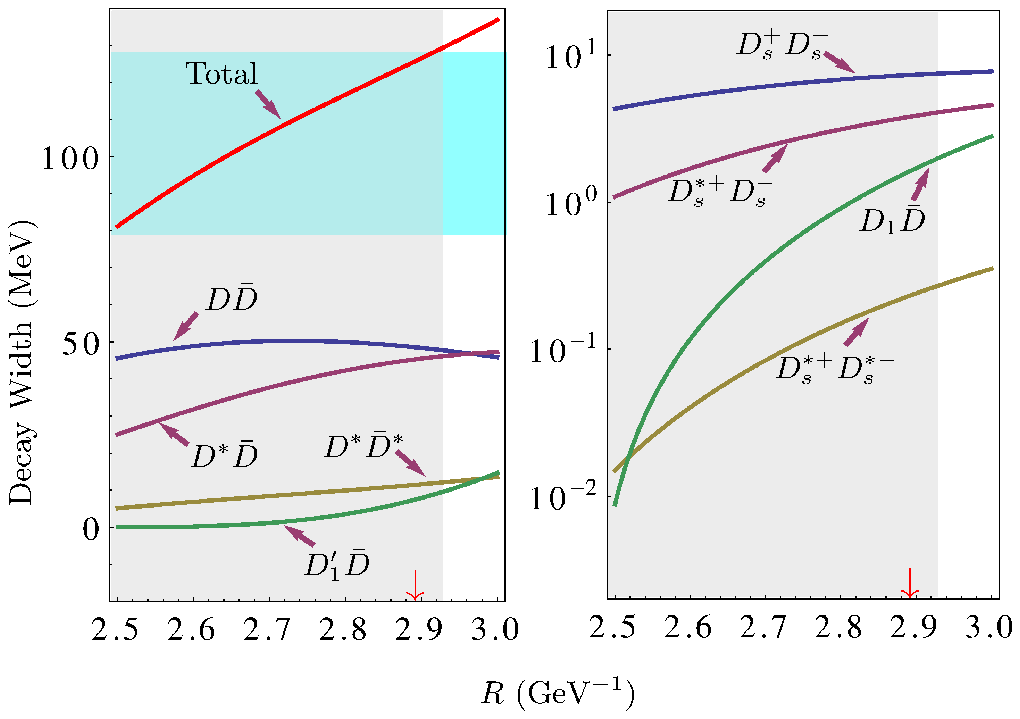}}
\caption{(Color online).  The same as Fig. \ref{Fig:psi12d} but for $\psi_1(3D)$\label{Fig:psi13d}}
\end{figure}

\begin{figure}[htb]
\vspace{0.25cm}
\scalebox{0.8}{\includegraphics{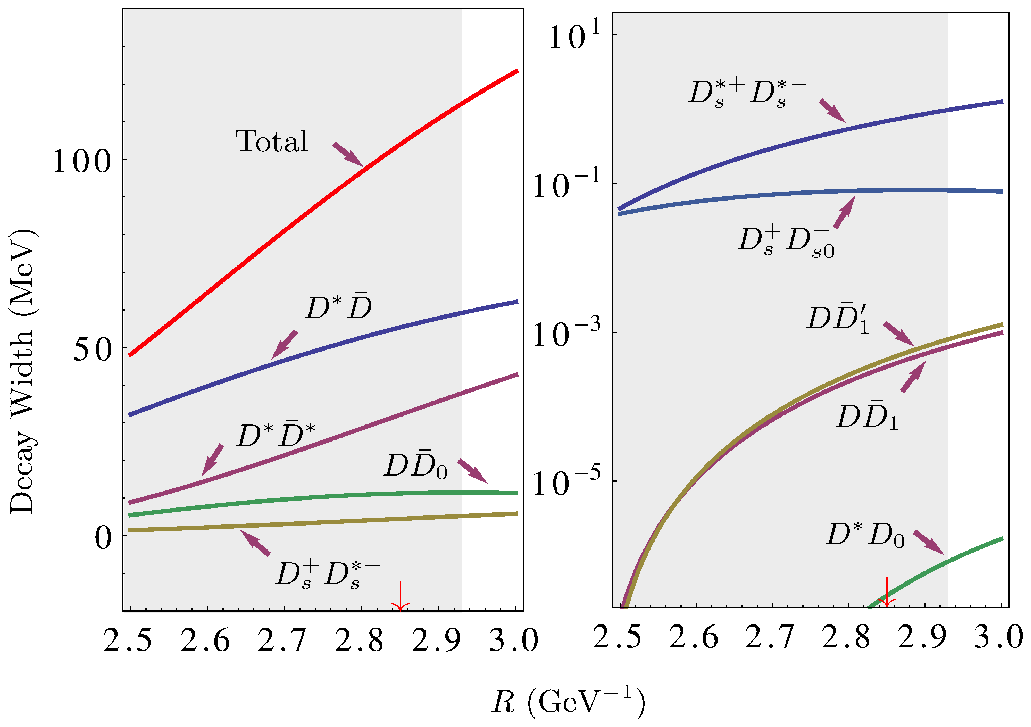}}
\caption{(Color online). Partial and total widths of $\eta_{c2}(3D)$. The light grey band is the $R$ range determined by the comparison of $Y(4320)$ total width with the experimental data, where $Y(4320)$ is assigned as $\psi_1(3D)$ charmonium. \label{Fig:etac23d}}
\end{figure}

\begin{figure}[htb]
\vspace{0.25cm}
\scalebox{0.8}{\includegraphics{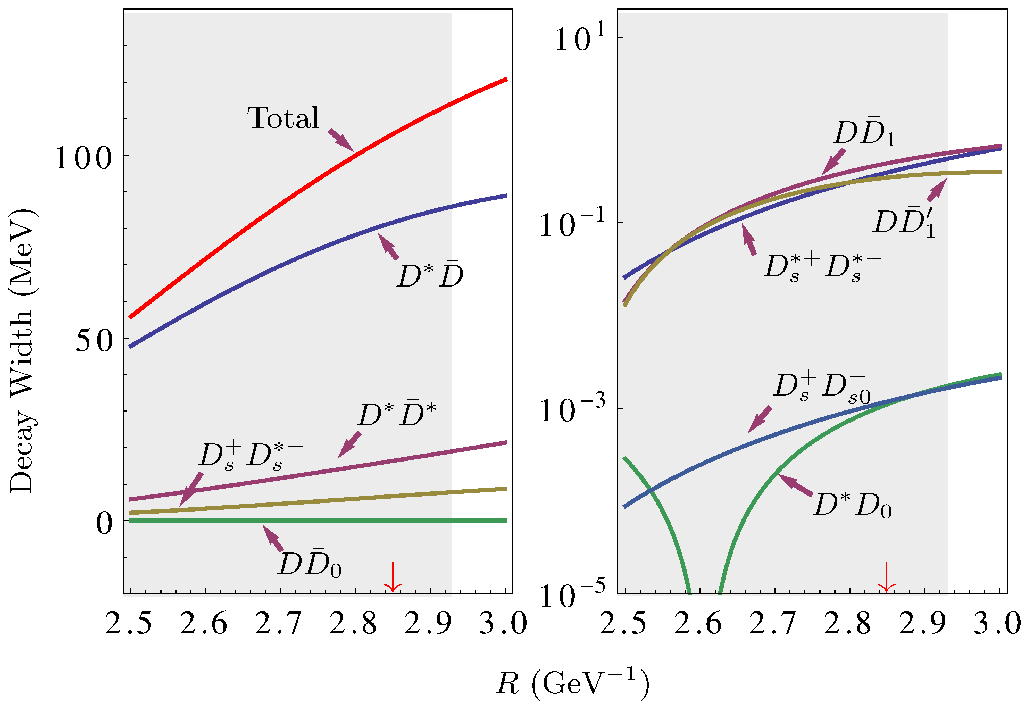}}
\caption{(Color online). The same as Fig. \ref{Fig:etac23d} but for $\psi_2(3D)$ charmonium. \label{Fig:psi23d}}
\end{figure}
\begin{figure}[htb]
\vspace{0.25cm}
\scalebox{0.8}{\includegraphics{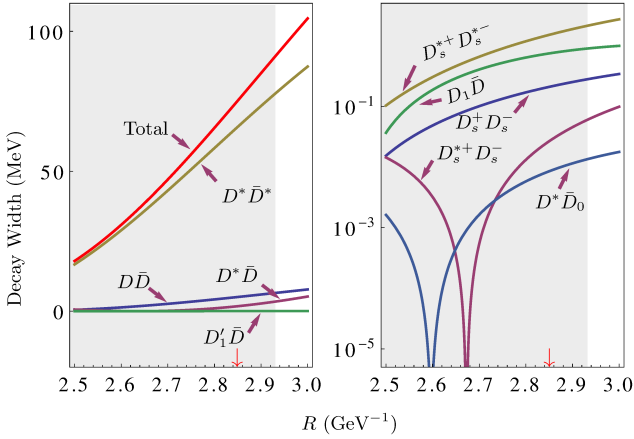}}
\caption{(Color online). The same as Fig. \ref{Fig:etac23d} but for $\psi_3(3D)$ charmonium. \label{Fig:psi33d}}
\end{figure}

The $R$ dependent total and partial widths of the open charm decays of $Y(4320)$ are presented in Fig. \ref{Fig:psi13d}, where $Y(4320)$ is assigned as $\psi(3^3D_1)$ charmonium. In the MGI model, the $R$ value of $3D$ charmonia is $2.85 \ \mathrm{GeV}^{-1}$ and with this $R$ value, the width of $\psi(3^3D_1)$ is $121.7$ MeV, which is consistent with one of $Y(4320)$, i.e., $(101.4^{+25.3}_{-19.7} \pm 10.2) \ \mathrm{MeV}$ \cite{Ablikim:2016qzw}.  To further check the $R$ dependence of the total width and partial width of $\psi(3^3D_1)$, we vary $R$ value from $2.5 \ \mathrm{GeV}^{-1}$ to $3.0 \ \mathrm{GeV}^{-1}$.  Our estimations indicate that when $R=2.50 \sim 2.92 \ \mathrm{GeV}^{-1}$ the evaluated total width are consistent with the measured one from the BESIII Collaboration \cite{Ablikim:2016qzw}. This $R$ value for $\psi(3^3D_1)$ is larger than the one of $\psi(4160)$ and $\psi(3770)$, which is consistent with our expectation. In this $R$ range, $\psi(3^3D_1)$ dominantly decays into $D\bar{D}$, $D^\ast \bar{D}$ and $D^\ast \bar{D}^\ast$.  The partial width of $\psi(3^3D_1) \to D\bar{D}$  weakly depend on the paramter $R$, and in the determined $R$ range, we find $\Gamma(\psi(3^3D_1)\to D\bar{D})= 45.2 \sim 48.0\ \mathrm{MeV}$. And in this $R$ range, the partial widths of $D^\ast \bar{D}$ and $D^\ast \bar{D}^\ast$ are estimated to be  $24.3 \sim 46.0 $ and $4.9 \sim 11.9 \ \mathrm{MeV}$,  respectively. The ratios of the partial widths of these dominant decay channels are predicted to be
\begin{eqnarray}
\Gamma(Y(4320)\to D \bar{D})\over \Gamma(Y(4320) \to D^\ast \bar{D}^\ast) &=&4.0 \sim 9.2 \nonumber\\
\Gamma(Y(4320)\to D^\ast \bar{D})\over\Gamma(Y(4320) \to D^\ast \bar{D}^\ast)&=&3.8 \sim 4.5, \nonumber
\end{eqnarray}
respectively.

As for $\eta_{c2}(3D)$, the total and partial widths depending on the model parameter $R$ are presented in Fig. \ref{Fig:etac23d}. In the $R$ range determined by $Y(4320)$, we find the total width of $\eta_{c2}(3D)$ is strongly dependent on the model parameter. In particular, the total width is estimated to be  $47 \sim 114$ MeV in this $R$ range.  Moreover, our estimation indicates that the $D^\ast \bar{D}$ and $D^\ast \bar{D}^\ast$ should be the dominant decay modes of $\eta_{c2}(3D)$ and the  partial widths ratio of these two modes is estimated to be,
\begin{eqnarray}
\frac{\Gamma(\eta_{c2} \to D^\ast \bar{D})}{ \Gamma(\eta_{c2} \to D^\ast \bar{D}^\ast)} =1.6 \sim 3.8.
\end{eqnarray}
The total and partial widths of $\psi_2(3D)$ are presented in Fig. \ref{Fig:psi23d}. The total width is estimated to be $54 \sim 113$ MeV and in the determined $R$ range, $D^\ast \bar{D}$ is the dominant decay modes, the branching ratio of this decay mode is $(76 \sim 86) \%$. As shown in Fig. \ref{Fig:psi33d},  the total width of $\psi_3(3D)$ is also strong dependent on the model parameter and predicted to be $17\sim 89$ MeV. In the determined $R$ range, the dominant decay mode is $D^\ast \bar{D}^\ast$, and its branching ratio is estimated to be $(86 \sim 93) \%$.

It should be noticed the measured width of $Y(4320)$ has a relative large uncertainty, thus in a large $R$ range our estimation could overlap with the measured data.  Then the predicted total and partial widths for $\eta_{c2}(3D),\ \psi_{2}(3D)$ and $\psi_3(3D)$ vary in a relative large range. However, the dominant decay modes and the partial widths ratios are weakly dependent on the model parameter, which are helpful for searching these missing $3D$ states.

\section{Summary}
The observations of the vector charmonium-like states in the $e^+ e^-$ annihilation processes make the states between $4.0$ and $ 4.5$ GeV overcrowed. Besides the higher excited $J/\psi$ state, these charmonium-like states could also be higher $\psi(^3D_1)$ states. Moreover, a $\psi_3(1D)$ candidate was observed in the $D\bar{D}$ invariant mass spectroscopy very recently by the LHCb Collaboration. These experimental measurements stimulate us to comb $D$ wave charmonium states. In the present work, by investigating the open charm decay behaviors, we evaluate the possibility of $Y(4320)$ as $\psi(3^3D_1)$ charmonium and then take $\psi(3770)$, $\psi(4160)$ and $Y(4320)$ as the scales to evaluate the open charm decays of other $1D$, $2D$ and $3D$ charmonium states.

Our estimations indicate that the total widths of $\psi(3770)$ and $\psi(4160)$ can be reproduced in a proper $R $ range, which are $R=1.60 \sim 1.76\ \mathrm{GeV}^{-1}$  and $1.82\sim 1.97)\ \mathrm{GeV}^{-1}$ for $\psi(3770)$ and $\psi(4160)$, respectively. As for $Y(4320)$, the estimated total width can overlap with the measured one when we take $R=2.50 \sim 2.92 \ \mathrm{GeV}^{-1}$, which indicates $Y(4320)$ can be a $\psi_1(3D)$ candidate. It should be notice that the $R$ range for $1D$  is very close to the one of estimated in MGI model, and for $3D$ charmonium, the $R$ value estimated in MGI model is consistent with the $R$ range determined by the width of $\psi(3^3D_1)$ state. However, as for $2D$ state, the $R$ value estimated in MGI model is much larger than the range determined by the width of $\psi(4160)$.

Taking $\psi(3770)$, $\psi(4160)$ and $\psi(4320)$ as the scale of $1D$, $2D$ and $3D$ charmonia, respectively, we can estimated the open charm decays of other $D$ wave charmonium states.  From our estimations, we find,
\begin{itemize}
\item $\psi(4382)$ could be assigned as $\psi_3(1D)$ state. The narrow width resulted from the high partial wave suppression and our estimated partial width $\psi_3(1D) \to D\bar{D}$ is safely under the measured width of $\psi(3842)$ and consistent with the theoretical estimations of other group.

\item As for $2D$ states, we predict the total widths of $\eta_{c2}(2D)$, $\psi_2(2D)$ and $\psi_3(2D)$ to be $48 \sim 64 $, $50\sim 52$ and $52 \sim 76$ MeV, respectively. We also find that the dominant decay modes of these three $D$ wave charmonia are $D^\ast \bar{D}$ and $D^\ast \bar{D}^\ast$, respectively. Furthermore, the partial widths ratios of these dominant channels are also predicted.

\item As for $3D$ states,  the predicted total widths of $\eta_{c2}(3D)$, $\psi_2(3D)$ and $\psi_3(3D)$ are in a relative large range  due to  the large uncertainty of $R$ determined by the width of $Y(4320)$. However, we find the predominant decay modes are $D^\ast\bar{D}$, $D^\ast\bar{D}$ and $D^\ast\bar{D}^\ast$ for $\eta_{c2}(3D)$, $\psi_2(3D)$ and $\psi_3(3D)$, respectively. Moreover,  some partial widths ratios are predicted, which are nonsensitive to the model parameter $R$.

\end{itemize}
The open charm decay channels are the important observation channels of higher charmonia since they are the dominant decay channels of these higher charmonia. The charmonia produced in hadronproduction process have more possible $J^{PC}$ quantum numbers. Thus,  hadronproduction  is one of most promising process of searching for the higher $D$ wave charmonia in the open charm mass spectroscopy. All the estimations in the present work could be helpful for searching for the missing $2D$ and $3D$ charmonia in the open charm decay channels in the further experimental measurements at LHCb.

\section*{Acknowledgement}
The authors would like to thank Jun-Zhang Wang for useful discussion. This project is supported by the National Natural Science Foundation of China under Grant No. 11775050, No. 11375240 and No. 11675228, Nature Science Foundation Projects of Qinghai Office of Science and Technology, No. 2017-ZJ-748, the Chunhui Plan of China’s Ministry of Education, No. Z2017054, the Natural Science Foundation of Jiangsu Province of China under contract No. BK20171349.

\appendix
\section{Partial Wave Amplitudes}
\label{Sec:App}.

\begin{table*}[htb]
\centering
\caption{The partial wave amplitude of the open charm decays for $^1D_2$ initial state. Here $\alpha=1/\sqrt{3}$ is the factor resulted from the flavor matrix element and $I^{M_{L_A},m}_{M_{L_B},M_{L_C}}$ is defined in Eq.~(\ref{Eq:I}).  \label{Tab:amp1}}
\renewcommand\arraystretch{1.35}
\begin{tabular}{p{4.0cm}<{\centering}p{12cm}<{\centering}}
\toprule[1pt]
Decay Channel  &  Amplitude\\
\midrule[1pt]
$^1D_2\to\  ^1S_0\  ^3S_1$ &  $\mathcal{M}^{11} = \frac{2\alpha}{5 \sqrt{3}} \sqrt{E_A E_B E_C} \gamma \left(  I^{00}_{00} -\sqrt{3} I^{1-1}_{00} \right) $ \\
 & $\mathcal{M}^{13} = \frac{\sqrt{2}\alpha}{15 } \sqrt{E_A E_B E_C} \gamma \left( -3 I^{00}_{00} -2\sqrt{3} I^{1-1}_{00} \right) $ \\
$^1D_2\to\  ^1S_0\  ^3S_1$ &  $\mathcal{M}^{11} = \frac{2\sqrt{2}\alpha}{15 } \sqrt{E_A E_B E_C} \gamma \left( \sqrt{3}  I^{00}_{00} -3 I^{1-1}_{00} \right) $ \\
 & $\mathcal{M}^{13} = -\frac{\sqrt{2}\alpha}{15 } \sqrt{E_A E_B E_C} \gamma \left(  3 I^{00}_{00} +2\sqrt{3} I^{1-1}_{00} \right) $ \\
$^1D_2\to\  ^1S_0\  ^3P_0$ & $\mathcal{M}^{02}=-\frac{\sqrt{2} \alpha}{3\sqrt{5}}  \sqrt{E_A E_B E_C} \gamma \left(  I^{00}_{00} +2  I^{01}_{01} \right) $ \\
$^1D_2\to\  ^1S_0\  P_1$ & $\mathcal{M}^{12}=\cos \theta\Big[ -\sqrt{\frac{2}{15}}  \alpha \sqrt{E_A E_B E_C} \gamma \left(  I^{1-1}_{00} +2  I^{10}_{01} \right)  \Big]$ \\
$^1D_2\to\  ^1S_0\  P_1^\prime$ & $\mathcal{M}^{12}=\sin \theta\Big[ -\sqrt{\frac{2}{15}}  \alpha \sqrt{E_A E_B E_C} \gamma \left(  I^{1-1}_{00} +2  I^{10}_{01} \right)  \Big]$ \\
$^1D_2\to\  ^3S_1\  ^3P_0$ & $\mathcal{M}^{12}=-\frac{2\alpha}{3\sqrt{5}}  \sqrt{E_A E_B E_C} \gamma \left(  I^{1-1}_{00} +2  I^{10}_{01} \right) $ \\
\bottomrule[1pt]
\end{tabular}
\end{table*}

\begin{table*}[htb]
\centering
\caption{The same as Table. \ref{Tab:amp1} but for $^3D_1$ initial state.  \label{Tab:amp2}}
\renewcommand\arraystretch{1.35}
\begin{tabular}{p{4.0cm}<{\centering}p{12cm}<{\centering}}
\toprule[1pt]
Decay Channel  &  Amplitude\\
\midrule[1pt]
$^3D_1\to\  ^1S_0\  ^1S_0$ &  $\mathcal{M}^{01} = \frac{2\alpha}{3 \sqrt{5}} \sqrt{E_A E_B E_C} \gamma \left(\sqrt{3} I^{1-1}_{00} -I^{00}_{00} \right) $ \\
$^3D_1\to  ^3S_1\  ^1S_0$ & $\mathcal{M}^{11} = \frac{2 \alpha}{\sqrt{30}} \sqrt{E_A E_B E_C} \gamma \left(\sqrt{3} I^{00}_{00} -3I^{1-1}_{00} \right) $ \\
$^3D_1\to  ^3S_1\  ^3S_1$ & $ \mathcal{M}^{01} = \frac{2\alpha}{3 \sqrt{15}} \sqrt{E_A E_B E_C} \gamma \left(I^{00}_{00} -\sqrt{3} I^{1-1}_{00} \right)$ \\
& $\mathcal{M}^{21} = \frac{2\alpha}{45} \sqrt{E_A E_B E_C} \gamma \left( 3I^{1-1}_{00} -\sqrt{3} I^{00}_{00} \right)$ \\
& $\mathcal{M}^{23} = \frac{2\sqrt{2} \alpha}{\sqrt{45}} \sqrt{E_A E_B E_C} \gamma \left(3 I^{00}_{00} +2 \sqrt{3} I^{1-1}_{00} \right)$ \\
$^3D_1\to  ^1S_0\  P_1^\prime$ &  $\mathcal{M}^{10}= \cos \theta   \Big[\frac{2\alpha}{9\sqrt{5}}    \sqrt{E_A E_B E_C} \gamma \Big( -\sqrt{3} I^{00}_{00} +\sqrt{3} I^{01}_{01}  +3I^{1-1}_{00} -3I^{10}_{01} +3\sqrt{2}I^{2-1}_{01}\Big) \Big]$\\
&$ \hspace{7.5mm} +\sin \theta   \Big[\frac{\alpha}{9} \sqrt{\frac{2}{5}}  \sqrt{E_A E_B E_C} \gamma \Big(  \sqrt{3} I^{00}_{00} -\sqrt{3} I^{01}_{01}  -3I^{1-1}_{00} +3I^{10}_{01} -3\sqrt{2}I^{2-1}_{01}\Big) \Big]$\\
& $\mathcal{M}^{12}=\cos \theta   \Big[ \frac{\alpha}{9\sqrt{5}}    \sqrt{E_A E_B E_C} \gamma \Big( 2 \sqrt{6} I^{00}_{00} +\sqrt{6} I^{01}_{01} -6 \sqrt{2}I^{1-1}_{00}-3\sqrt{2}I^{10}_{01}+6 I^{2-1}_{01}  \Big) \Big] $ \\
&$ \hspace{12mm} +\sin \theta   \Big[\frac{\alpha}{9\sqrt{5}}    \sqrt{E_A E_B E_C} \gamma \Big( \sqrt{3} I^{00}_{00} +5\sqrt{3} I^{01}_{01} -3 \sqrt{2}I^{1-1}_{00}-6\sqrt{2}I^{10}_{01}-3\sqrt{2} I^{2-1}_{01}  \Big) \Big]$\\
$^3D_1\to \  ^1S_0\  P_1$ &  $\mathcal{M}^{10}= -\sin \theta   \Big[\frac{2\alpha}{9\sqrt{5}}    \sqrt{E_A E_B E_C} \gamma \Big( -\sqrt{3} I^{00}_{00} +\sqrt{3} I^{01}_{01}  +3I^{1-1}_{00} -3I^{10}_{01} +3\sqrt{2}I^{2-1}_{01}\Big) \Big]$\\
&$ \hspace{7.5mm} +\cos \theta   \Big[\frac{\alpha}{9} \sqrt{\frac{2}{5}}  \sqrt{E_A E_B E_C} \gamma \Big(  \sqrt{3} I^{00}_{00} -\sqrt{3} I^{01}_{01}  -3I^{1-1}_{00} +3I^{10}_{01} -3\sqrt{2}I^{2-1}_{01}\Big) \Big]$\\
& $\mathcal{M}^{12}=-\sin \theta   \Big[ \frac{\alpha}{9\sqrt{5}}    \sqrt{E_A E_B E_C} \gamma \Big( 2 \sqrt{6} I^{00}_{00} +\sqrt{6} I^{01}_{01} -6 \sqrt{2}I^{1-1}_{00}-3\sqrt{2}I^{10}_{01}+6 I^{2-1}_{01}  \Big) \Big] $ \\
&$ \hspace{12mm} +\cos \theta   \Big[\frac{\alpha}{9\sqrt{5}}    \sqrt{E_A E_B E_C} \gamma \Big( \sqrt{3} I^{00}_{00} +5\sqrt{3} I^{01}_{01} -3 \sqrt{2}I^{1-1}_{00}-6\sqrt{2}I^{10}_{01}-3\sqrt{2} I^{2-1}_{01}  \Big) \Big]$\\
$^3D_1\to\ ^1S_0\  ^3P_2$ & $\mathcal{M}^{22} = \frac{ \alpha}{3\sqrt{5}} \sqrt{E_A E_B E_C} \gamma \left(- I^{00}_{00} + I^{01}_{01}+ \sqrt{3} I^{1-1}_{00} -\sqrt{6} I^{2-1}_{01} \right)$   \\
$^3D_1\to\ ^3S_1\  ^3P_0$ & $\mathcal{M}^{1,2}=\frac{\alpha}{3} \sqrt{\frac{2}{5}} \sqrt{E_A E_B E_C} \gamma \left(I^{00}_{00} +2 I^{01}_{01}+ \sqrt{3} I^{1-1}_{00} + I^{10}_{01} \right)$\\
\bottomrule[1pt]
\end{tabular}
\end{table*}

\begin{table*}[htb]
\centering
\caption{The same as Table. \ref{Tab:amp1} but for $^3D_2$ initial state.  \label{Tab:amp3}}
\renewcommand\arraystretch{1.35}
\begin{tabular}{p{4.0cm}<{\centering}p{12cm}<{\centering}}
\toprule[1pt]
Decay Channel  &  Amplitude\\
\midrule[1pt]
$^3D_2\to\  ^1S_0\  ^3S_1$ &  $\mathcal{M}^{11} = \frac{\sqrt{2}\alpha}{5} \sqrt{E_A E_B E_C} \gamma \left( I^{00}_{00} - \sqrt{3}I^{1-1}_{00} \right) $ \\
& $ \mathcal{M}^{13} =\frac{2\alpha}{15} \sqrt{E_A E_B E_C} \gamma \left( \sqrt{3} I^{00}_{00} +2I^{1-1}_{00} \right) $ \\
$^3D_2\to\  ^3S_1\  ^3S_1$ &  $\mathcal{M}^{21} = \frac{\sqrt{2}\alpha}{15} \sqrt{E_A E_B E_C} \gamma \left( 3 \sqrt{2} I^{1-1}_{00} - \sqrt{6}I^{00}_{00} \right) $ \\
& $ \mathcal{M}^{23} =\frac{4\alpha}{15} \sqrt{E_A E_B E_C} \gamma \left( \sqrt{3} I^{00}_{00} +2I^{1-1}_{00} \right) $ \\
$^3D_2\to\  ^1S_0\  ^3P_0$ &  $\mathcal{M}^{02} = \frac{2\alpha}{3\sqrt{5}} \sqrt{E_A E_B E_C} \gamma \left( \sqrt{3} I^{1-1}_{00} +2 I^{1-1}_{00} \right) $ \\
$^3D_2\to\  ^1S_0\  ^3P_0$ &  $\mathcal{M}^{02} = \frac{2\alpha}{3\sqrt{5}} \sqrt{E_A E_B E_C} \gamma \left( \sqrt{3} I^{1-1}_{00} +2 I^{1-1}_{00} \right) $ \\
$^3D_2\to\  ^1S_0\  P_1$ & $\mathcal{M}^{12} = -\sin \theta \Big[  \frac{\sqrt{2}\alpha}{3\sqrt{5}} \sqrt{E_A E_B E_C} \gamma \left( \sqrt{3} I^{01}_{01}- I^{10}_{01}-\sqrt{2} I^{2-1}_{01} \right)  \Big]   $\\
& $ \hspace{12mm} +\cos\theta \Big[ \frac{\alpha}{3\sqrt{5}} \sqrt{E_A E_B E_C} \gamma \left( \sqrt{3} I^{00}_{00}+ \sqrt{3}I^{01}_{01}- I^{1-1}_{00} + \sqrt{2} I^{2-1}_{01} \right) \Big]$\\
$^3D_2\to\  ^1S_0\  P_1^\prime$ & $\mathcal{M}^{12} = \cos \theta \Big[  \frac{\sqrt{2}\alpha}{3\sqrt{5}} \sqrt{E_A E_B E_C} \gamma \left( \sqrt{3} I^{01}_{01}- I^{10}_{01}-\sqrt{2} I^{2-1}_{01} \right)  \Big]   $\\
& $ \hspace{12mm} +\sin\theta \Big[ \frac{\alpha}{3\sqrt{5}} \sqrt{E_A E_B E_C} \gamma \left( \sqrt{3} I^{00}_{00}+ \sqrt{3}I^{01}_{01}- I^{1-1}_{00} + \sqrt{2} I^{2-1}_{01} \right) \Big]$\\
$^3D_2\to\  ^3S_1\  ^3P_0$ &  $\mathcal{M}^{12} = \frac{\sqrt{2}\alpha}{9\sqrt{5}} \sqrt{E_A E_B E_C} \gamma \left(3 I^{00}_{00} +6I^{01}_{01} +\sqrt{3} I^{1-1}_{00} +I^{10}_{01}\right) $ \\
\bottomrule[1pt]
\end{tabular}
\end{table*}

\begin{table*}[htb]
\centering
\caption{The same as Table. \ref{Tab:amp1} but for $^3D_3$ initial state.  \label{Tab:amp4}}
\renewcommand\arraystretch{1.35}
\begin{tabular}{p{4.0cm}<{\centering}p{12cm}<{\centering}}
\toprule[1pt]
Decay Channel  &  Amplitude\\
\midrule[1pt]
$^3D_3\to\  ^1S_0\  ^1S_0$ &  $\mathcal{M}^{03} = \frac{\sqrt{2}\alpha}{3\sqrt{35}} \sqrt{E_A E_B E_C} \gamma \left( 3 I^{00}_{00} +2 \sqrt{3}I^{1-1}_{00} \right) $ \\
$^3D_3\to\  ^1S_0\  ^3S_1$ &  $ \mathcal{M}^{13} =\frac{2\sqrt{2}\alpha}{3\sqrt{35}} \sqrt{E_A E_B E_C} \gamma \left( \sqrt{3} I^{00}_{00} +2I^{1-1}_{00} \right) $ \\
$^3D_3\to\  ^3S_1\  ^3S_1$ &  $\mathcal{M}^{03} =-\frac{\sqrt{2}\alpha}{3\sqrt{35}} \sqrt{E_A E_B E_C} \gamma \left( \sqrt{3} I^{00}_{00} +2I^{1-1}_{00} \right) $\\
&  $\mathcal{M}^{21} =\frac{2\sqrt{2} \alpha}{15} \sqrt{E_A E_B E_C} \gamma \left( 3\sqrt{2} I^{1-1}_{00} -\sqrt{6}I^{00}_{00} \right) $\\
&  $\mathcal{M}^{23} =\frac{4 \alpha}{15\sqrt{7}} \sqrt{E_A E_B E_C} \gamma \left( 3 I^{00}_{00} +2\sqrt{3}I^{1-1}_{00} \right) $\\
$^3D_3\to\  ^1S_0\  P_1$ & $\mathcal{M}^{12}= -\sin \theta\Big[ \frac{\sqrt{2} \alpha}{21 \sqrt{5}} \sqrt{E_A E_B E_C} \gamma \left( 3\sqrt{3} I^{00}_{00} +4\sqrt{3}I^{01}_{01}  + 6I^{1-1}_{00}+8 I^{10}_{01}+2\sqrt{2}I^{2-1}_{01}\right) \Big] $\\
&\hspace{12mm}  $\cos \theta \Big[\frac{\sqrt{2} \alpha}{21 \sqrt{5}} \sqrt{E_A E_B E_C} \gamma \left(2\sqrt{6} I^{00}_{00} +5\sqrt{6}I^{01}_{01}  + 4 \sqrt{2} I^{1-1}_{00}+3 \sqrt{2} I^{10}_{01}-2 I^{2-1}_{01}\right)
\Big] $ \\
& $\mathcal{M}^{14}= -\sin \theta\Big[ \frac{\sqrt{2} \alpha}{21 \sqrt{5}} \sqrt{E_A E_B E_C} \gamma \left( -6 I^{00}_{00} +6 I^{01}_{01}  -4 \sqrt{3}I^{1-1}_{00}+4\sqrt{3} I^{10}_{01}+\sqrt{6}I^{2-1}_{01}\right) \Big] $\\
&\hspace{12mm}  $\cos \theta \Big[\frac{\sqrt{2} \alpha}{105} \sqrt{E_A E_B E_C} \gamma \left(3\sqrt{10} I^{00}_{00} -3\sqrt{10}I^{01}_{01}  +2 \sqrt{30} I^{1-1}_{00}-2 \sqrt{30} I^{10}_{01}-\sqrt{15} I^{2-1}_{01}\right)
\Big] $ \\
$^3D_3\to\  ^1S_0\  P_1^\prime$ & $\mathcal{M}^{12}= \cos \theta\Big[ \frac{\sqrt{2} \alpha}{21 \sqrt{5}} \sqrt{E_A E_B E_C} \gamma \left( 3\sqrt{3} I^{00}_{00} +4\sqrt{3}I^{01}_{01}  + 6I^{1-1}_{00}+8 I^{10}_{01}+2\sqrt{2}I^{2-1}_{01}\right) \Big] $\\
&\hspace{12mm}  $\sin \theta \Big[\frac{\sqrt{2} \alpha}{21 \sqrt{5}} \sqrt{E_A E_B E_C} \gamma \left(2\sqrt{6} I^{00}_{00} +5\sqrt{6}I^{01}_{01}  + 4 \sqrt{2} I^{1-1}_{00}+3 \sqrt{2} I^{10}_{01}-2 I^{2-1}_{01}\right)
\Big] $ \\
& $\mathcal{M}^{14}= \cos \theta\Big[ \frac{\sqrt{2} \alpha}{21 \sqrt{5}} \sqrt{E_A E_B E_C} \gamma \left( -6 I^{00}_{00} +6 I^{01}_{01}  -4 \sqrt{3}I^{1-1}_{00}+4\sqrt{3} I^{10}_{01}+\sqrt{6}I^{2-1}_{01}\right) \Big] $\\
&\hspace{12mm}  $\sin \theta \Big[\frac{\sqrt{2} \alpha}{105} \sqrt{E_A E_B E_C} \gamma \left(3\sqrt{10} I^{00}_{00} -3\sqrt{10}I^{01}_{01}  +2 \sqrt{30} I^{1-1}_{00}-2 \sqrt{30} I^{10}_{01}-\sqrt{15} I^{2-1}_{01}\right)
\Big] $ \\
$^3D_3\to\  ^1S_0\  ^3P_2$ &  $\mathcal{M}^{22} =\frac{2 \sqrt{2}\alpha}{21\sqrt{5}} \sqrt{E_A E_B E_C} \gamma \left( \sqrt{3} I^{00}_{00} -\sqrt{3}I^{01}_{01} +2 I^{1-1}_{00}+5 I^{10}_{01} +3\sqrt{2}I^{2-1}_{01} \right) $\\
&  $\mathcal{M}^{24} =\frac{\sqrt{2} \alpha}{21} \sqrt{E_A E_B E_C} \gamma \left( -\sqrt{6} I^{00}_{00} +\sqrt{6}I^{01}_{01} -2\sqrt{2} I^{1-1}_{00}+2\sqrt{2} I^{10}_{01} +I^{2-1}_{01} \right)  $\\
$^3D_3\to\  ^3S_1\  ^3P_0$ &  $\mathcal{M}^{12} =\frac{\sqrt{2}\alpha}{9\sqrt{5}} \sqrt{E_A E_B E_C} \gamma \left(3 I^{00}_{00} +6I^{01}_{01} +2 I^{1-1}_{00}-2\sqrt{3} I^{1-1}_{00} +I^{10}_{01} \right) $\\
\bottomrule[1pt]
\end{tabular}
\end{table*}

\end{document}